\documentclass[11pt]{article}
\usepackage{a4wide}
\usepackage{graphicx}           

\usepackage{color}

\usepackage{amsmath,amssymb}
\newcommand{\df}{\text{d}}

\usepackage{epsf}

\newcommand{\GeV}{\ensuremath{\,\text{GeV} }}

\newcommand{\al}[1]{\begin{align}#1\end{align}}
\newcommand{\paren}[1]{\left(#1\right)}

\newcommand{\ab}[1]{\left|#1\right|}

\usepackage{epsf}

\newcommand{\R}{\mathcal{R}}

\begin{document}

\title{
\bf \Large Higgs inflation still alive\vskip 0.5cm
}
\author{
Yuta~Hamada\thanks{E-mail: \tt hamada@gauge.scphys.kyoto-u.ac.jp},
Hikaru~Kawai\thanks{E-mail: \tt hkawai@gauge.scphys.kyoto-u.ac.jp}, 
Kin-ya~Oda\thanks{E-mail: \tt odakin@phys.sci.osaka-u.ac.jp}, and
Seong~Chan~Park\thanks{E-mail: \tt s.park@skku.edu}
\bigskip\\
\it \normalsize
$^{*\dagger}$ Department of Physics, Kyoto University, Kyoto 606-8502, Japan\\
\it \normalsize
$^\ddag$ Department of Physics, Osaka University, Osaka 560-0043, Japan\\
\it \normalsize
$^{\S}$ Department of Physics, Sungkyunkwan University, Suwon 440-746, Korea\smallskip
}
\date{\today}


\maketitle

\abstract{\noindent\normalsize
The observed value of the Higgs mass indicates that the Higgs potential becomes small and flat at the scale around $10^{17}\,\text{GeV}$.
Having this fact in mind, we reconsider the Higgs inflation scenario proposed by Bezrukov and Shaposhnikov.
It turns out that the non-minimal coupling $\xi$ of the Higgs-squared to the Ricci scalar can be smaller than ten.
For example, $\xi=7$ corresponds to the tensor-to-scalar ratio $r\simeq0.2$, which is consistent with the recent observation by BICEP2.
}

\newpage

\normalsize

\begin{figure}
\begin{center}
\includegraphics[width=0.5\textwidth]{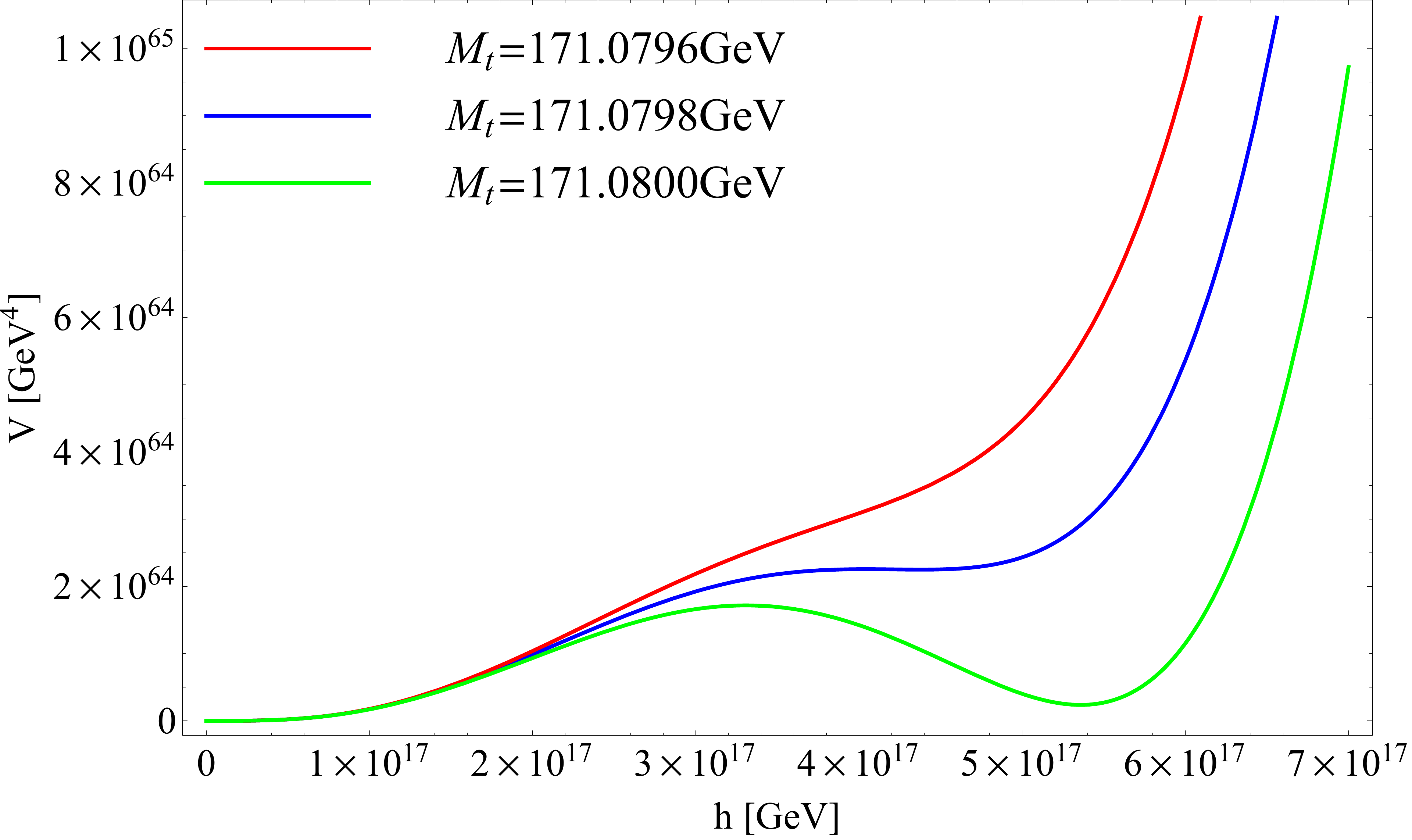}
\caption{Standard model Higgs potential for the Higgs mass 125.6\,GeV.}
\label{xi=0 figure}
\end{center}
\end{figure}

The observed value of the Higgs mass $125.9\pm0.4\GeV$~\cite{PDG2014} indicates that the Standard Model (SM) Higgs potential becomes small and flat at the scale around $10^{17}\,\text{GeV}$; see e.g.~\cite{Holthausen:2011aa,Bezrukov:2012sa,Degrassi:2012ry,Alekhin:2012py,Masina:2012tz,Hamada:2012bp,Jegerlehner:2013cta,Buttazzo:2013uya} for latest analyses.\footnote{
It is an intriguing fact that the bare Higgs mass also becomes small at the same scale~\cite{Hamada:2012bp,Hamada:2013cta,Masina:2013wja}; see also Refs.~\cite{Alsarhi:1991ji,Jones:2013aua,Haba:2013lga}. The running Higgs mass after the subtraction of the quadratic divergence is considered e.g.\ in Ref.~\cite{Bian:2013xra}.
}
See Fig.~\ref{xi=0 figure} for the Higgs potential around that scale for various values of the top quark mass~\cite{Hamada:2013cta}.
We see that by tuning the top quark mass, we can make the first derivative at the inflection point arbitrarily small as shown by the blue (center) line.
Note that the required tuning of the top quark mass is rather strict.
The values of $M_t$ are given to show the amount of tuning and should not be taken literally.\footnote{
The latest combined result for the top quark mass is $173.34\pm0.76\GeV$~\cite{ATLAS:2014wva}.
Note that there can be a discrepancy between the pole mass $M_t$ and the one measured at the hadron colliders; see e.g.\ Refs.~\cite{Alekhin:2012py,Hamada:2013cta}.
The latter is obtained as an invariant mass of the color singlet final states, whereas the former is a pole of a colored quark. At the hadron colliders, the observed $t\bar t$ pair is dominantly color octet, and there may be discrepancy of order 1-2 GeV in drawing extra lines to make the singlet final states. We thank Yukinari Sumino on this point. See also Ref.~\cite{Horiguchi:2013wra}.}
There are several arguments that this tuning is required by a principle such as the multiple point principle~\cite{Froggatt:1995rt,Froggatt:2001pa,Nielsen:2012pu}, the maximum entropy principle~\cite{Kawai:2011qb,Kawai:2013wwa}, the classical conformality~\cite{Meissner:2006zh,Foot:2007iy,Meissner:2007xv,Iso:2009ss,Iso:2009nw,Aoki:2012xs,Iso:2012jn,Hashimoto:2013hta,Hashimoto:2014ela}, and the asymptotic safety~\cite{Shaposhnikov:2009pv}.

It is known that this inflection point cannot be used to achieve a successful inflation~{\cite{Isidori:2007vm,Hamada:2013mya}}.\footnote{
See e.g.\ Refs.~\cite{Allahverdi:2006iq,Imai:2013zia} for attempts of the inflection point inflation.
}
Slow-roll condition $\ab{\eta_V}\lesssim1$ restricts the field value to be very close to the inflection point.
To earn a sufficient $e$-folding $N_*\simeq60$ within this range of $\varphi_*$, the first derivative at the inflection point must be very small, and hence cannot yield the right amount of the amplitude $A_s\propto V_*/\epsilon_V$ at $\varphi_*$.

In Ref.~{\cite{Hamada:2013mya}}, we have discussed a possibility that a new physics, such as string theory, modifies the Higgs potential above the scale $\Lambda\sim 10^{17}\GeV$.
In this Letter, we pursue another possibility that the non-minimal coupling of the Higgs-squared to the Ricci scalar, $\xi\varphi^2\R$, leads to a successful inflection point inflation.

The main differences from the ordinary Higgs inflation scenario~\cite{Bezrukov:2007ep,Park:2008hz,Bezrukov:2009db,Bezrukov:2010jz,Salvio:2013rja}
are the following two points:\footnote{
For the other attempts, see Refs.~\cite{Kamada:2010qe,Kamada:2012se,Kamada:2013bia,Nakayama:2014koa}.
See also Refs.~\cite{Buchbinder:1992rb,Martin:2013tda}.}
\begin{itemize}
\item The $e$-folding is earned in passing the inflection point, and hence the relation $\epsilon_V\sim 1/N_*^2$ no longer holds.
Therefore, the scalar-to-tensor ratio $r=16\epsilon_V$ can be sizable to match the recent BICEP2 result~\cite{BICEP2}:
\al{
r	&=	0.2^{+0.07}_{-0.05}
}
at the 68\% CL.
\item $\xi$ can be smaller than ten, since the Higgs quartic coupling $\lambda$ is small at $\varphi_*$ due to the tuning mentioned above.
\end{itemize}

We start from the same Lagrangian as the ordinary Higgs inflation~\cite{Bezrukov:2007ep,Bezrukov:2009db,Bezrukov:2010jz}.\footnote{
Here we are treating the Higgs inflation as a single-field model, as in Refs.~\cite{Bezrukov:2007ep,Bezrukov:2009db,Bezrukov:2010jz}.
Even when the Nambu-Goldstone bosons are explicitly included,
the multi-field version of Higgs inflation rapidly evolves as an effectively single-field model due to the dynamics of the system's evolution~\cite{Greenwood:2012aj,Kaiser:2013sna}, rather than by tacitly assuming unitary gauge.
The analysis in Ref.~\cite{Greenwood:2012aj} did not incorporate quantum corrections as the present paper does, but nonetheless, based on Ref.~\cite{Greenwood:2012aj} as well as Ref.~\cite{Kaiser:2013sna}, it should be clear that the single-field (dynamical) attractor behavior holds whenever the non-minimal coupling is sufficiently large.
}
The potential in the Einstein frame can be obtained from the effective potential
\al{
V(\varphi)
	&=	{\lambda(\varphi)\over4}\varphi^4
		\label{potential}
}
in the flat space, by setting $\varphi=\varphi_h$ with
\al{
\varphi_h
	:=	{h\over\sqrt{1+\xi h^2/M_P^2}},
		\label{phi_h}
}
where $h$ is the Higgs field in the Jordan frame.\footnote{
This choice corresponds to the prescription I in Ref.~\cite{Bezrukov:2009db},
 which minimizes the one-loop logarithmic correction to the effective potential in the Einstein frame.
}

The running coupling $\lambda(\mu)$ has a minimum at $\mu_\text{min}\sim 10^{17\text{--}18}\GeV$, depending on the Higgs mass~\cite{Holthausen:2011aa,Bezrukov:2012sa,Degrassi:2012ry,Alekhin:2012py,Masina:2012tz,Hamada:2012bp,Jegerlehner:2013cta,Buttazzo:2013uya}.\footnote{
The Higgs quartic coupling grows above the minimum due to the contribution of the growing $U(1)_Y$ coupling.
Qualitatively, the position and height of the minimum depend on the Higgs and top masses, respectively.
}
Around the minimum, $\lambda(\mu)$ can be expanded as
\al{
\lambda(\mu)
	&=	\lambda_\text{min}
		+{\beta_2\over\paren{16\pi^2}^2}\paren{\ln{\mu\over\mu_\text{min}}}^2
		+{\beta_3\over\paren{16\pi^2}^3}\paren{\ln{\mu\over\mu_\text{min}}}^3
		+\cdots,
}
where $\beta_2\simeq0.6$ in the SM~\cite{Hamada:2013mya}. The term proportional to $\beta_3$ and higher are small in the region of our interest, and we will neglect them hereafter.
The value of $\lambda_\text{min}$ depends on the top quark mass,
and we can set it arbitrarily small by tuning the top quark mass within the current experimental bound. 

For the potential $V(\varphi)$ to be monotonically increasing around the inflection point, it is necessary and sufficient that
\al{
\lambda_\text{min}
	\geq	\lambda_c
	:=		{\beta_2\over\paren{64\pi^2}^2}\sim 10^{-6}.
}
The equality holds when the potential has a plateau.
That is, when we put $\lambda_\text{min}=\lambda_c$, the point $\varphi_\text{inflection}=e^{-1/4}\mu_\text{min}\simeq0.8\mu_\text{min}$ becomes a saddle point with vanishing first and second derivatives.\footnote{
There appears another inflection point at $e^{-11/12}\mu_\text{min}\simeq 0.4\mu_\text{min}$ too.
} 

We set the value of $\lambda_\text{min}$ slightly larger than $\lambda_c$ to realize an inflection point inflation, while keeping the potential above $\varphi_\text{inflection}$ sufficiently small by the introduction of $\xi$ in order to evade the problem described above. 
The three cases $\lambda>\lambda_c$, $\lambda=\lambda_c$, and $\lambda<\lambda_c$ corresponds to the red (upper), blue (middle), and green (lower) curves in Fig.~\ref{xi=0 figure}, respectively.
An important point here is that the value of $\varphi_h$ in Eq.~\eqref{phi_h} is saturated to $M_P/\sqrt{\xi}$ for large values of $h$ ($\gg M_P/\sqrt{\xi}$), and therefore the potential does not grow rapidly. In order for this saturation to work to avoid too large $\eta_V$, we need $\varphi_\text{inflection}\sim M_P/\sqrt{\xi}$, that is, $\xi\sim M_P^2/\mu_\text{min}^2$.

\begin{figure}
\begin{center}
\hfill
\includegraphics[width=0.4\textwidth]{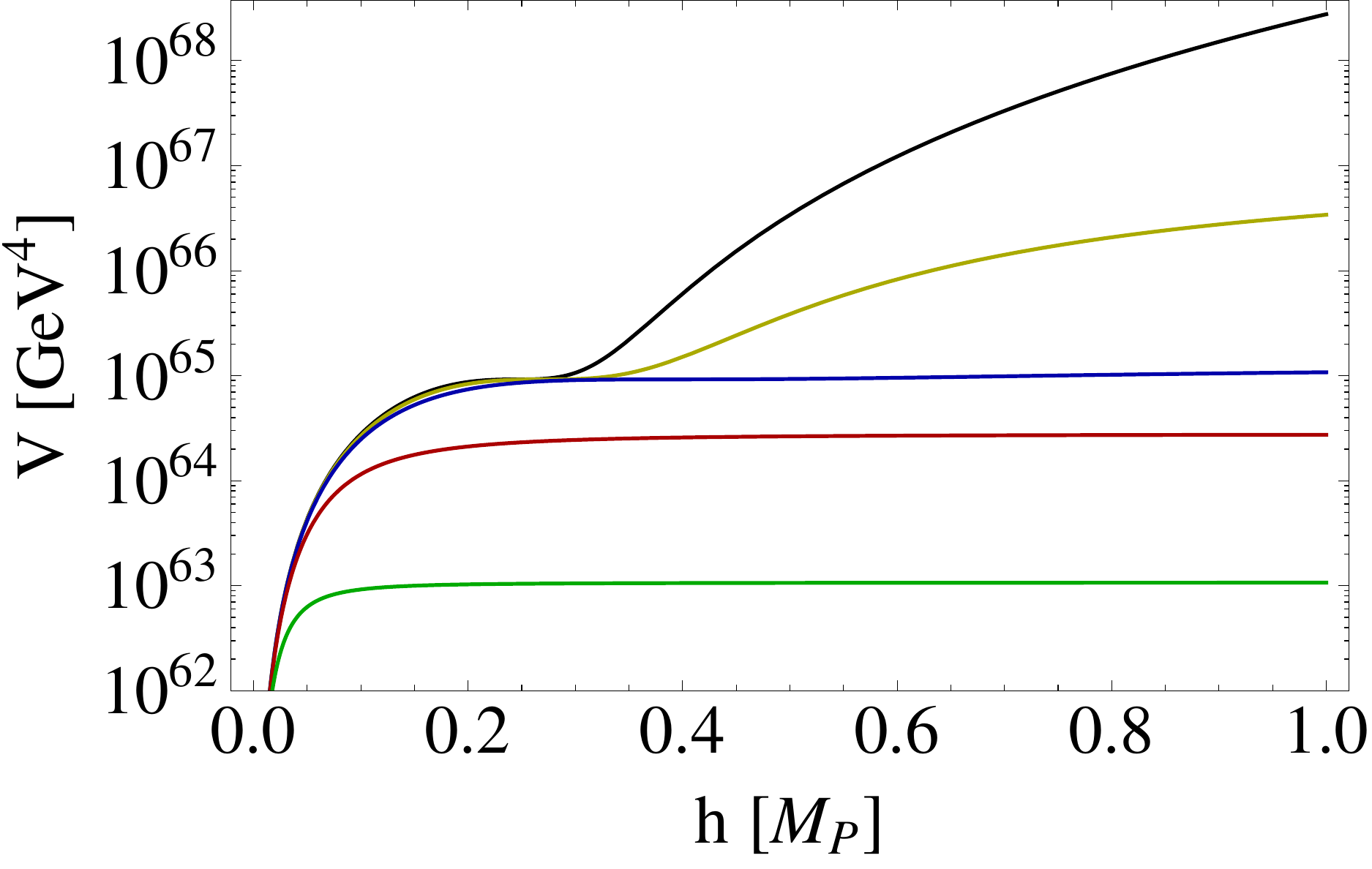}\hfill
\includegraphics[width=0.4\textwidth]{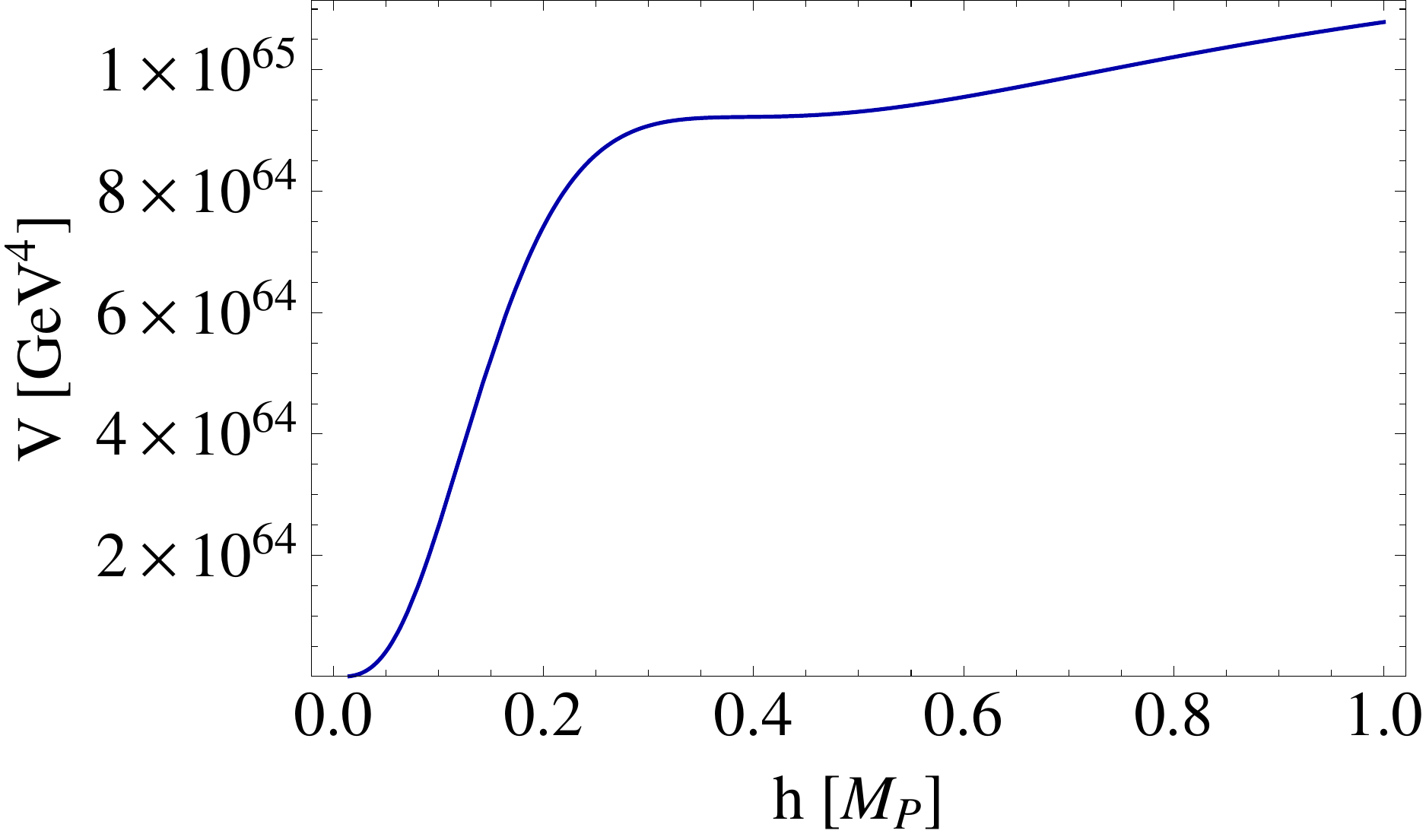}
\hfill\mbox{}
\caption{Left: Inflaton potential for $\xi=0,3,10,100$, and 1000 from above to below in log-linear plot.
Right: the same for $\xi=10$ in linear-linear plot.}
\label{potential figure}
\end{center}
\end{figure}

As concrete examples, we show our results for several benchmark points with the parameter choice $\xi=0,3,10,100$, and 1000 with $\lambda_\text{min}=1.01\lambda_c$, $\beta_2=0.6$, and $\mu_\text{min}=M_P/\sqrt{10}$
in the left panel in Fig.~\ref{potential figure}; the same figure is drawn in linear plot for $\xi=10$ in the right panel.

\begin{figure}
\begin{center}
\hfill
\includegraphics[width=0.4\textwidth]{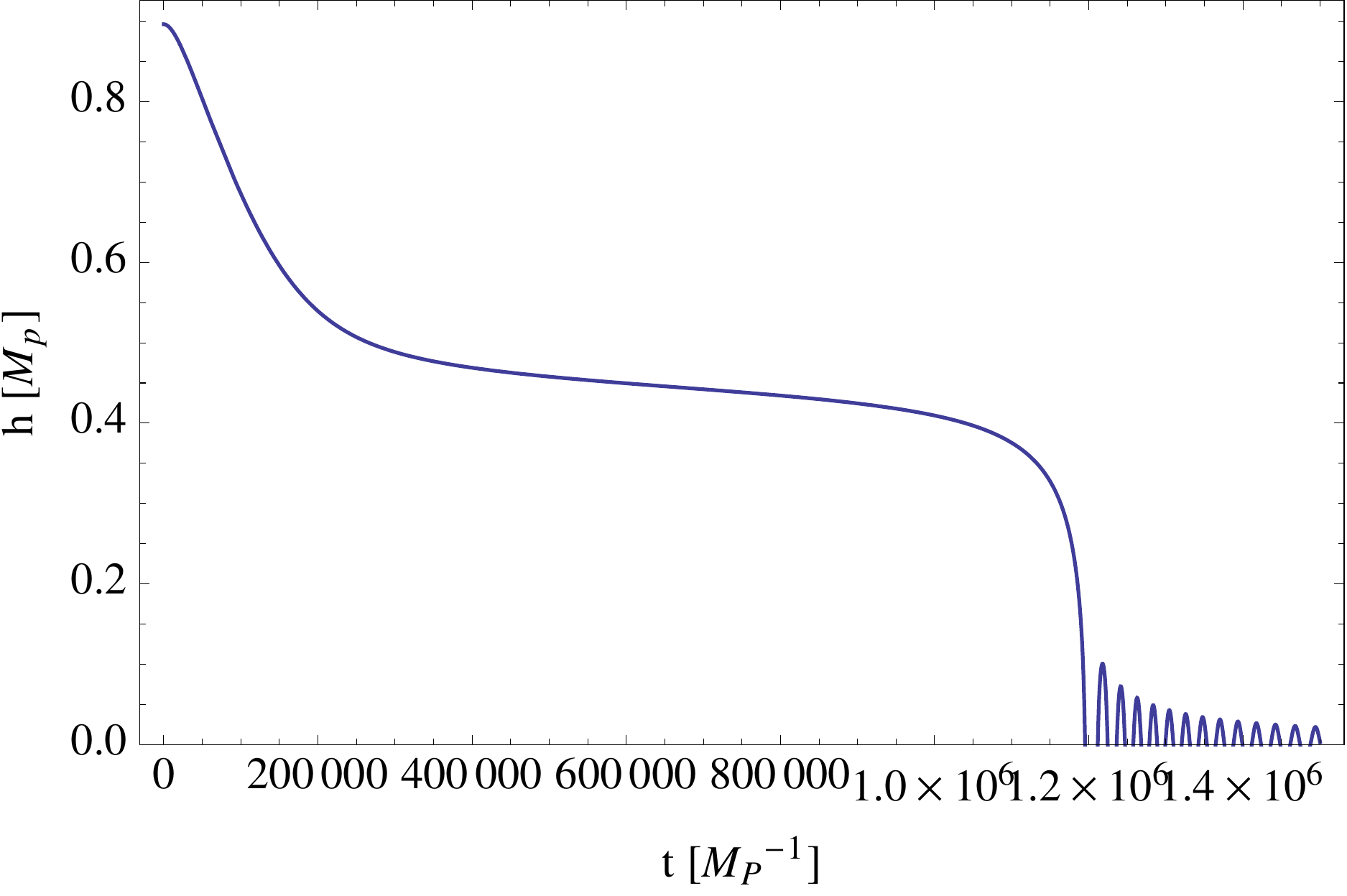}
\hfill\mbox{}
\caption{$h$ vs $t$ in Einstein frame in Planck unit.}
\label{evolution figure}
\end{center}
\end{figure}

To fit the cosmological data, we can e.g.\ take
$h_*=0.896M_P$, $\lambda_\text{min}=1.01\lambda_c$, $\mu_\text{min}=0.37M_P$, $\xi=7$ to get $r=16\epsilon_V(h_*)=0.19$, $N_*=58$, $V(\varphi_{h_*})/\epsilon_V(h_*)=5.0\times10^{-7}$ and $n_s(h_*)=0.955$, where
\al{
\epsilon_V
	&=	{M_P^2\over2V(\varphi_h)^2}\paren{{\df h\over \df\chi}{\df V(\varphi_h)\over\df h}}^2,&
\eta_V
	&=	{M_P^2\over V(\varphi_h)}{\df h\over \df\chi}{\df\over\df h}\paren{{\df h\over \df\chi}{\df V(\varphi_h)\over\df h}},
}
with
\al{
{\df\chi\over\df h}
	&=	{\sqrt{1+\xi\paren{1+6\xi}h^2/M_P^2}\over1+\xi h^2/M_P^2}.
}
For the same parameters, the Einstein-frame time evolution of the Higgs field $h$ is plotted in Fig.~\ref{evolution figure}.
We see that substantial time is spent around the inflection point.

Once the tensor-to-scalar ratio is fixed to be $r\simeq0.2$, the slow-roll parameter becomes $\epsilon_V(h_*)\simeq0.013$, and the amplitude $A_s\propto V(\varphi_{h_*})/\epsilon_V(h_*)$ fixes the potential height $V(\varphi_{h_*})^{1/4}\simeq2\times10^{16}\GeV$.
The potential height is determined in our case to be $V(\varphi_{h_*})\simeq\lambda(\varphi_{h_*})M_P^4/\xi^2$, which is the same as the Higgs inflation.
The difference is the value of $\lambda(\varphi_{h_*})\simeq\lambda_\text{min}\simeq\lambda_c\sim 10^{-6}$ that allows us to take $\xi\lesssim 10$.

In this Letter, we have matched the renormalization scale in the Einstein frame, as in Eq.~\eqref{phi_h}. If we instead match it in the Jordan frame,\footnote{
It is argued in Ref.~\cite{Bezrukov:2013fka} that these two choices correspond to different theories. The choice $\varphi=h$ corresponds to the prescription II in Ref.~\cite{Bezrukov:2009db}, which minimizes the one-loop logarithmic correction in the Jordan frame.
We note however that there is a criticism on this choice~\cite{George:2013iia}.
Whether calculations involving quantum corrections are or are not frame-dependent is a point of ongoing research, and we leave it open for further discussion.
}
i.e.\ if we set $\varphi=h$ in Eq.~\eqref{potential}, we obtain the chaotic inflation at $h\gg M_P/\sqrt{\xi}$.
In this region, the canonically normalized field is $\hat\chi\simeq\sqrt{6}M_P\ln{hM_P\over\sqrt{\xi}\mu_\text{min}^2}$ in the Einstein frame. The potential for $\hat\chi$ becomes quadratic:
\al{
V\simeq
	{\lambda_\text{min}M_P^4\over4\xi^2}
	+{1\over2}{\beta_2M_P^2\over48\xi^2\paren{16\pi^2}^2}\hat\chi^2.
}
We see that by taking $\xi\sim100$, we get the right amount of the inflaton mass $\sim10^{13}\GeV$.

Finally we comment on the unitarity issue in the Higgs inflation due to the large non-minimal coupling, which requires a new physics above the scale $\Lambda\sim M_P/\xi$ in order to cure the scattering being strongly coupled on the electroweak vacuum ($\varphi\ll M_P/\sqrt{\xi}$)~\cite{Burgess:2009ea,Barbon:2009ya,Burgess:2010zq,Giudice:2010ka,Burgess:2014lza}.\footnote{
We note that the cutoff scale depends on the background field value $\varphi$, and for $\varphi\gtrsim M_P/\sqrt{\xi}$, it becomes $\Lambda\sim\sqrt{M_P^2+\xi\varphi^2}$~\cite{Bezrukov:2010jz}. That is, inflation takes place below the cutoff scale.
}
It is an implicit assumption of the Higgs inflation that such an extension does not affect the result qualitatively, that is, the Wilson coefficients of the higher order terms are sufficiently smaller than $1/\xi$. Note that our $\xi$ is greatly reduced from the value $\xi\sim10^5$ in the ordinary scenario.

\subsection*{Note added:}
There appeared in the same day on arXiv an article treating similar subject~\cite{Cook:2014dga}, which is consistent with our result.\footnote{
In Ref.~\cite{Cook:2014dga}, the result of Ref.~\cite{Allison:2013uaa} in the prescription II is used.
}

\section*{Acknowledgement}
K.O.\ and S.C.P.\ thank Takuya Kakuda and Jinsu Kim for useful discussions.
We thank Yukinari Sumino for the useful comments.
The work of Y.H.\ is partly supported by a Grant-in-Aid from the Japan Society for the Promotion of Science (JSPS) Fellows No.~25.1107.
S.C.P.\ is supported by Basic Science Research Program through the National Research Foundation of Korea funded by the Ministry of Education, Science and Technology (2011-0010294) and (2011-0029758) and (NRF-2013R1A1A2064120).
The work of K.O.\ is in part supported by the Grant-in-Aid for Scientific Research Nos.~23104009, 20244028, and 23740192. 

\bibliographystyle{TitleAndArxiv}
\bibliography{Higgs_letter}
\end{document}